%% file: BESIII_PLB_31542.tex
\documentclass[preprint,3p]{elsarticle}
\usepackage{graphics}
\usepackage{amssymb}
\usepackage{amsthm}
\usepackage[mathlines]{lineno}
\usepackage{graphicx}
\usepackage{units}
\usepackage{url}
\usepackage{amsmath}
\usepackage{amsfonts}
\usepackage{bm}
\usepackage{textcomp}
\usepackage{subfigure}
\usepackage{multicol}
\usepackage{verbatim}
\usepackage[colorlinks,linkcolor=blue,citecolor=blue]{hyperref}

\journal{Physics Letters B}

\begin{document}

\begin{frontmatter}

\title{{\bf
\boldmath Measurement of the branching fraction for $\psi(3770) \to
\gamma\chi_{c0}$}}

\input{BESIII_authors}

\begin{abstract}
By analyzing a data set of 2.92~fb$^{-1}$ of $e^+e^-$ collision data taken at $\sqrt s= 3.773~\rm GeV$ and
106.41$\times 10^{6}$ $\psi(3686)$ decays taken at $\sqrt s= 3.686~\rm GeV$
with the BESIII detector at the BEPCII collider,
we measure the branching fraction and the partial decay width for
$\psi(3770)\to\gamma\chi_{c0}$ to be
${\mathcal B}(\psi(3770)\to\gamma\chi_{c0})=(6.88\pm0.28\pm0.67)\times 10^{-3}$ and
$\Gamma[\psi(3770)\to\gamma\chi_{c0}]=(187\pm8\pm19)~\rm keV$, respectively.
These are the most precise measurements to date.
\end{abstract}


\end{frontmatter}


\section{Introduction}
Transitions between charmonium states can be used to shed light on various aspects of
Quantum Chromodynamics~(QCD), the theory of the strong interactions,
in both the perturbative and non-perturbative regimes
\cite{charmonium}.
The $\psi(3770)$ resonance is the lowest-mass charmonium state lying
above the production threshold of open-charm $D\bar D$ pairs.
It is assumed to be the $1^3D_1$ $c\bar c$
state with a small $2^3S_1$ admixture. Based on this $S$-$D$ mixing
model, predictions have been made
\cite{prd64_094002,prd44_3562,prd69_094019,prd72_054026,new_cite} for the
partial widths of the $\psi(3770)$ electric-dipole ($E1$) radiative
transitions. These predictions vary over a large range depending on
the underlying model assumptions. One of the largest variations in predictions
is for the partial width of $\psi(3770)\to \gamma\chi_{c0}$, with
predictions ranging from 213~keV to 523~keV. A precise
measurement of the partial width of
$\psi(3770)\to\gamma\chi_{c0}$ provides a stringent test of the
various theoretical approaches, thereby providing a better
understanding of $\psi(3770)$ decays.

In 2006, the CLEO Collaboration reported the first observation of $\psi(3770)\to\gamma\chi_{c0/1}$
and measured the partial widths~\cite{prd74_031106,prl96_182002}. A comparison between their results
and predictions of traditional theory models \cite{prd64_094002,prd44_3562,prd69_094019,prd72_054026}
indicates that relativistic and
coupled-channel effects are necessary ingredients to describe the data.
A similar conclusion has been drawn in $\psi(3686)\to\gamma\chi_{cJ}$ decays~\cite{psip_g}.
The results of CLEO were normalized to the cross section of $\psi(3770)\to D\bar D$
to obtain the total number of $\psi(3770)$ decays, which assumed the contribution of
$\psi(3770)\to$ non-$D\bar D$ decays is negligible \cite{prl96_092002}.
Recently, the BESIII Collaboration presented an improved measurement of
$\psi(3770)\to\gamma\chi_{c1}$ \cite{bes3_psipp_gchi1}.

In this Letter, we report on an alternative and complementary measurement
of the branching fraction and partial width of $\psi(3770)\to\gamma\chi_{c0}$ using
$\chi_{c0}\to 2(\pi^+\pi^-)$, $K^+K^-\pi^+\pi^-$, $3(\pi^+\pi^-)$ and $K^+K^-$ decays.
The results of our measurements are obtained by taking the
relative strength with respect to the well-known $\psi(3686)$
radiative $E1$ transition~\cite{pdg2014}.
In this way,
the measurement will not depend on knowledge of the $\chi_{cJ}$ branching fractions to light hadron final states,
which have large uncertainties~\cite{prd74_031106}.
This measurement forms an independent and more precise benchmark that can be compared to the predictions of various theoretical models.

\section{BESIII detector and Monte Carlo simulation}
In this work, we use 2.92 fb$^{-1}$ of $e^+e^-$ collision data taken at
$\sqrt s= 3.773~\rm GeV$ \cite{jll_bes3_lum}, and 106.41 $\times 10^{6}$ $\psi(3686)$ decays taken at
$\sqrt s= 3.686~\rm GeV$ \cite{ref:num_psip} with the BESIII detector.
These are labeled the $\psi(3770)$ and $\psi(3686)$ data samples, respectively, throughout this Letter.

The BESIII detector \cite{bes3} has a geometrical acceptance of 93\% of 4$\pi$ and
consists of four main components.
In the following, we describe each detector component starting from the innermost
(closest to the interaction region) to the most outside layer.
The inner three components are immersed in the $1\;\unit{T}$ magnetic field of a
superconducting solenoid.
First, a small-cell, helium-based main drift chamber (MDC) with 43 layers
provides charged particle tracking and measurement of ionization energy loss ($dE/dx$).
The average single wire resolution is 135 $\rm \mu m$, and the momentum resolution for 1~GeV
electrons in a $1\,\unit{T}$ magnetic field is 0.5\%.
The next detector after the MDC is a time-of-flight system (TOF) used for particle
identification. It is composed of a barrel part made of two layers of
88 plastic scintillators, each with 5 cm thickness and 2.4 m length;
and two endcaps, each with 96 fan-shaped plastic scintillators of 5 cm
thickness. The time resolution is 80~ps in the barrel,
and 110~ps in the endcaps,
corresponding to a $K/\pi$ separation better than $2\sigma$ for momenta up to about 1.0 GeV.
The third detector component is an electromagnetic calorimeter (EMC) made of 6240 CsI(Tl) crystals
arranged in a cylindrical shape (barrel) plus two endcaps.
For 1.0 GeV photons, the energy resolution is 2.5\% in the barrel and 5\% in the endcaps,
and the position resolution is 6 mm in the barrel and 9 mm in the endcaps.
Outside the EMC, a muon chamber system (MUC) is incorporated in the
return iron of the superconducting magnet. It is made of 1272 m$^2$ of resistive plate chambers
arranged in 9 layers in the barrel and 8 layers in the endcaps.
The position resolution is about 2~cm.

A GEANT4 \cite{geant4} based Monte Carlo (MC) simulation software package, which includes the geometric
description of the detector and the detector response, is used to determine
the detection efficiency of the signal process and to estimate the potential peaking backgrounds.
Signal MC samples of $\psi(3686)/\psi(3770)\to \gamma\chi_{cJ}$
are generated
with the angular distribution that corresponds to an $E1$
transition, and the  $\chi_{cJ}$ decays to light hadron final states
are generated according to a phase-space model.  Particle decays are
modeled using EvtGen \cite{evtgen}, while the initial production is
handled by the MC generator KKMC \cite{kkmc}, in which both
initial state radiation (ISR) effects \cite{isr} and
final state radiation (FSR) effects \cite{photons} are considered.
For the background studies of $\psi(3686)$ decays,
106$\times 10^{6}$ MC events of generic decays $\psi(3686)\to \text{anything}$ are produced
at $\sqrt s=3.686$ GeV.
For the background studies of $\psi(3770)$ decays,
MC samples of
$\psi(3770) \to D^0\bar D^0$, $\psi(3770) \to D^+D^-$, $\psi(3770) \to$
non-$D\bar D$ decays, ISR production of $\psi(3686)$ and $J/\psi$,
QED, and $q\bar q$ continuum processes
are produced at $\sqrt s=3.773$ GeV.
The known decay modes of the $J/\psi$, $\psi(3686)$ and $\psi(3770)$
are generated with branching fractions taken from the PDG \cite{pdg2014},
and the remaining events are generated with Lundcharm \cite{lundcharm}.

\section{Analysis}
\label{sec:evtsel}

To select candidate events for
$\psi(3686)/\psi(3770)\to\gamma\chi_{cJ}$ with $\chi_{cJ}\to$
$2(\pi^+\pi^-)$/$K^+K^-\pi^+\pi^-$/$3(\pi^+\pi^-)$/ $K^+K^-$, we require
at least 4/4/6/2 charged tracks to be reconstructed in the
MDC, respectively.
All charged tracks used in this analysis are required to be within
a polar-angle ($\theta$) range of $|\rm{cos~\theta}|<0.93$.
It is required that all charged tracks originate from the
interaction region defined by $|V_z|<10$ cm and $|V_{xy}|<1$ cm,
where $|V_z|$ and $|V_{xy}|$ are the distances of closest approach of the charged track to
the collision point in the beam direction and
in the plane perpendicular to the beam, respectively.

Charged particles are identified by confidence levels for kaon and
pion hypotheses calculated using $dE/dx$ and TOF measurements. To
effectively separate pions and kaons, a track is identified as a
pion (or kaon) only if the confidence level for the pion (or kaon)
hypothesis is larger than the confidence level for the kaon (or
pion) hypothesis.

Photons are selected by exploiting the information from the EMC.
It is required that the shower time be within 700 ns of the event start time
and the shower energy be greater than 25 (50) MeV
in the barrel (endcap) region defined by $|\cos\theta|<0.80$ ($0.86<|\cos\theta|<0.92$).
Here, $\theta$ is the photon polar angle with respect to the beam direction.

In the selection of $\gamma 2(\pi^+\pi^-)$,
background events from radiative Bhabha events
in which at least two radiative photons are produced and
one of them converts into an $e^+e^-$ pair
are suppressed by requiring the opening angle of any
$\pi^+\pi^-$ combination be larger than $10^{\circ}$.
For the selection of $\gamma K^+K^-$,
the background events of $e^+e^-\to \gamma e^+e^-$ are suppressed
by requiring $E_{\rm EMC}<1$ GeV and $E_{\rm EMC}/p_{\rm MDC}<0.8$ for each charged kaon,
where $E_{\rm EMC}$ and $p_{\rm MDC}$ are
the energy deposited in the EMC and the momentum measured by the MDC,
respectively.

In each event, there may be several different charged and/or neutral
track combinations which satisfy the selection criteria for each
light hadron final state. Each combination is subjected to
a $4C$ kinematic fit for the hypotheses of
$\psi(3686)/\psi(3770)\to$ $\gamma 2(\pi^+\pi^-)$, $\gamma
K^+K^-\pi^+\pi^-$, $\gamma 3(\pi^+\pi^-)$ and $\gamma K^+K^-$.
For each final state, if more than one combination satisfies the selection
criteria, only the combination with the least $\chi^2_{4C}$ is
retained, where $\chi^2_{4C}$ is the chi-square of the $4C$ kinematic fit.
The final states with $\chi^2_{4C}<25$ are kept for further analysis.

\begin{figure}[htbp]
\begin{center}
\includegraphics*[width=9.0cm]
{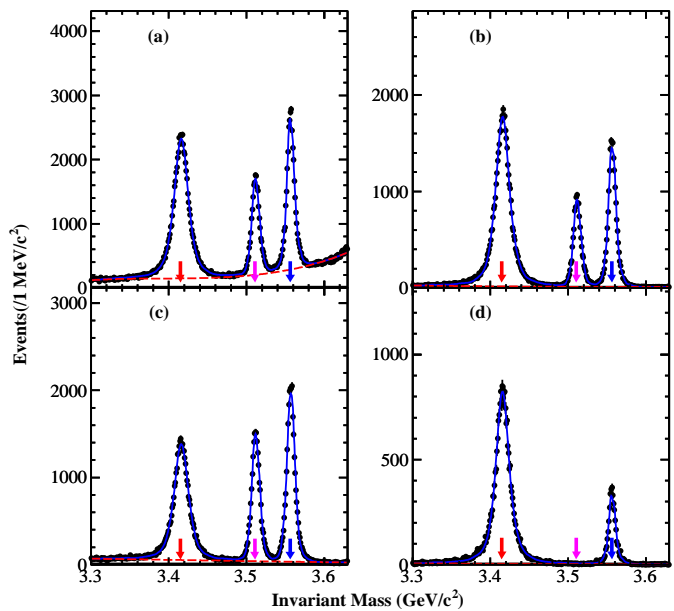}
 \caption{\small
Invariant mass spectra of the
(a) $2(\pi^+\pi^-)$,
(b) $K^+K^-\pi^+\pi^-$,
(c) $3(\pi^+\pi^-)$ and
(d) $K^+K^-$ combinations
for the $\psi(3686)$ data.
The dots with error bars are for data and
the blue solid lines are the fit results.
The red dashed lines are the fitted backgrounds.
The red, pink and blue arrows show the $\chi_{c0}$,
$\chi_{c1}$ and $\chi_{c2}$ nominal masses, respectively.
}
\label{fig:psi3686}
\end{center}
\end{figure}

To identify the $\chi_{cJ}$ decays, we examine the
invariant mass spectra of the light hadron final states.
Figure \ref{fig:psi3686} shows the corresponding mass spectra for the $\psi(3686)$ data,
in which clear $\chi_{c0}$, $\chi_{c1}$ and $\chi_{c2}$ signals are observed.
Since the $\chi_{c1}$ cannot decay into two
pseudoscalar mesons because of spin-parity conservation,
the $\chi_{c1}$ signal cannot be observed in the $K^+K^-$ invariant mass
spectrum.
By fitting these spectra separately,
we obtain the numbers of $\chi_{cJ}$ observed from the
$\psi(3686)$ data, $N_{\psi(3686)}$,
which are summarized in Table~\ref{tab:num_data}. In
the fits, the $\chi_{cJ}$ signals are described
by the MC simulated line-shapes convoluted by Gaussian
functions for the resolution. Backgrounds in the four channels are
described by 3/3/3/1-parameter polynomial functions.
The parameters of the convoluted Gaussian functions
and the Chebychev polynomial functions are all free.

\begin{figure}[htbp]
\begin{center}
\includegraphics*[width=9.0cm]
{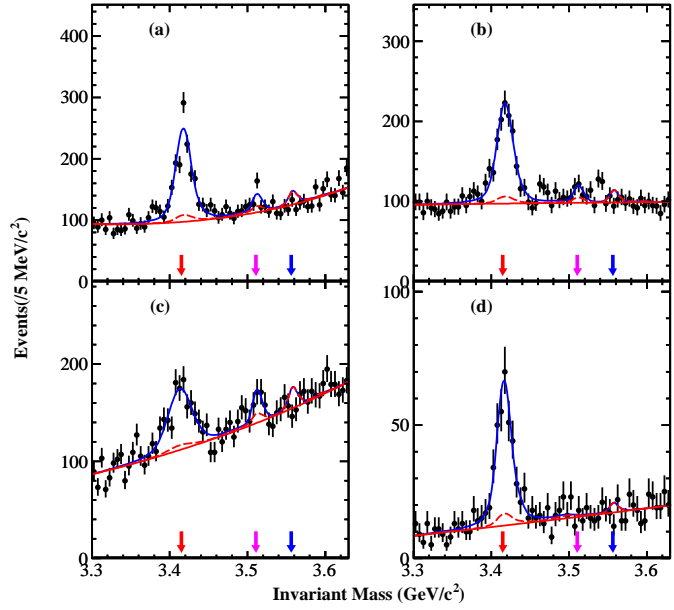} \caption{\small
Invariant mass spectra of the
(a) $2(\pi^+\pi^-)$,
(b) $K^+K^-\pi^+\pi^-$,
(c) $3(\pi^+\pi^-)$ and
(d) $K^+K^-$ combinations
for the $\psi(3770)$ data.
The dots with error bars are data and
the blue solid lines are the fit results.
the red solid lines are the fitted combinatorial backgrounds.
The red dashed lines are the sums of the peaking and fitted combinatorial backgrounds.
The red, pink and blue arrows show the $\chi_{c0}$,
$\chi_{c1}$ and $\chi_{c2}$ nominal masses, respectively.
}
\label{fig:psi3770}
\end{center}
\end{figure}

Figure \ref{fig:psi3770} shows the corresponding mass spectra for the $\psi(3770)$ data,
in which clear peaks can be observed for the $\chi_{c0}$ decays.
Fitting to these spectra similarly, we
obtain the number of $\chi_{cJ}(J=0,1)$ decays observed from the
$\psi(3770)$ data, $N_{\psi(3770)}$, which are summarized in Table~\ref{tab:num_data}.
Due to the limited statistics, the decay $\psi(3770)\to \gamma \chi_{c2}$
is not further considered in this analysis.
The means and widths of the convoluted Gaussian
functions for the $\chi_{c0}$ signals are left free. For the $\chi_{c1}$,
the mean and width of the convoluted Gaussian functions are
fixed at the values taken from the fits to the $\psi(3686)$ data.
Backgrounds in the four channels are
described by 6/2/6/2-parameter polynomial functions.

The background events from $e^+e^-\to(\gamma_{\rm
ISR})\psi(3686)$ produced near $\sqrt s = 3.773$ GeV have the same
event topologies as those from $\psi(3770)$ decays and are
indistinguishable from $\psi(3770)$ decays.
In the fits to the $\psi(3770)$ data, the size and line-shape of
such backgrounds are fixed according to MC simulations,
with the numbers of background events being determined by
\begin{equation}
N^{\rm b}_{\chi_{cJ}}
=\sigma^{\chi_{cJ},LH,\rm obs}_{\psi(3686)}
\cdot \mathcal{L}_{\psi(3770)}
\cdot \eta,
\label{eq:bk_psip}
\end{equation}
where
$\mathcal{L}_{\psi(3770)}$ is the integrated luminosity of the $\psi(3770)$ data,
$\sigma^{\chi_{cJ},LH,\rm obs}_{\psi(3686)}$ is the observed cross section
of $e^+e^-\to\psi(3686)\to \gamma\chi_{cJ}$ with $\chi_{cJ}\to LH$,
in which $LH$ denotes $2(\pi^+\pi^-)$, $K^+K^-\pi^+\pi^-$, $3(\pi^+\pi^-)$ and $K^+K^-$.
In this work, we assume that
there is no other effect affecting the $\psi(3686)$ and
$\psi(3770)$ production in the energy range from 3.73 to 3.89 GeV.
The variable $\eta$ represents the rate of misidentifying $\psi(3686)$ decays
as $\psi(3770)$ decays, which is obtained by analyzing $1.5\times 10^{6}$ MC events
of $\psi(3686)\to\gamma\chi_{cJ}$ with
$\chi_{cJ}\to LH$ generated at $\sqrt s=3.773$ GeV.
The observed cross section for $\psi(3686)\to\gamma\chi_{cJ}$
with $\chi_{cJ}\to LH$ at a center-of-mass energy of $\sqrt s$ is given by

\begin{equation}\label{Eq_XSEC_psip_obs}
      \sigma_{\psi(3686)}^{\chi_{cJ},LH,\rm obs} =
\int \sigma^{\chi_{cJ},LH}_{\psi(3686)}(s^\prime)
f(s^\prime)
F(x, s) G(s,s^{\prime\prime}) ds^{\prime\prime} dx.
\end{equation}
where $s^\prime \equiv s(1-x)$ is the square of the actual center-of-mass energy of
the $e^+e^-$ after radiating photon(s), $x$ is the fraction of the radiative energy to the beam energy;
$f(s^\prime)$ is the phase space factor, $({E_\gamma}(s^\prime)/{E_\gamma^0} )^3,$
in which
$E_\gamma(s^\prime)$ and $E_\gamma^0$ are the photon energies in $\psi(3686)\to\gamma\chi_{cJ}$
transition at $\sqrt {s^\prime}$
and at the $\psi(3686)$ mass, respectively;
$F(x,s)$ is the sampling function describing the radiative photon energy fraction $x$ at
$\sqrt s$~\cite{isr};
$G(s,s^{\prime\prime})$ is a Gaussian function describing the distribution of the collision
energy with an energy spread $\sigma_E=1.37$ MeV as achieved at BEPCII;
$\sigma^{\chi_{cJ},LH}_{\psi(3686)}(s^\prime)$ is the cross section
described by the Breit-Wigner function

\begin{equation}\label{Eq_XSEC_psip_B}
\sigma_{\psi(3686)}^{\chi_{cJ},LH}(s^\prime)
    = \frac{12\pi\Gamma^{ee}_{\psi(3686)}\Gamma^{\rm tot}_{\psi(3686)}
  B_{\psi(3686)}^{\chi_{cJ},LH}
 }
      {(s^{\prime2}-M_{\psi(3686)}^2)^2 + (\Gamma^{\rm tot}_{\psi(3686)}M_{\psi(3686)})^2},
\end{equation}
in which
$\Gamma^{ee}_{\psi(3686)}$ and $\Gamma^{\rm tot}_{\psi(3686)}$ are,
respectively,
the leptonic width and total width of the $\psi(3686)$,
$M_{\psi(3686)}$ is the $\psi(3686)$ mass,
${\mathcal B}_{\psi(3686)}^{\chi_{cJ},LH}$
is the combined branching fraction of $\psi(3686)\to\gamma\chi_{cJ}$
with $\chi_{cJ}\to LH$.
Here, the upper limit of $x$ is set at $1-m^2_{\chi_{cJ}}/s$,
where $m_{\chi_{cJ}}$ is the $\chi_{cJ}$ nominal mass.
We determine the branching fraction ${\mathcal B}_{\psi(3686)}^{\chi_{cJ},LH}$
by dividing the number of $\chi_{cJ}$
decays of $\psi(3686)$ by the total number of $\psi(3686)$ and
by the corresponding efficiency obtained in this work.
The rates $\eta$ of misidentifying $\psi(3686)\to\gamma\chi_{c0/1/2}$
as $\psi(3770)\to\gamma\chi_{c0/1/2}$
are estimated to be
4.72/6.40/7.60$\times 10^{-4}$,
4.40/6.27/7.57$\times 10^{-4}$,
3.53/4.95/6.14$\times 10^{-4}$ and
6.56/-/11.02$\times 10^{-4}$
for $\chi_{c0/1/2}\to 2(\pi^+\pi^-)$, $K^+K^-\pi^+\pi^-$, $3(\pi^+\pi^-)$ and $K^+K^-$,
respectively.
These lead to the number of background events from
$e^+e^-\to(\gamma_{\rm ISR})\psi(3686)$ to be
$90.6\pm3.4$/$37.5\pm1.4$/$76.5\pm2.9$,
$70.0\pm2.7$/$23.5\pm0.9$/$51.0\pm1.9$,
$56.6\pm2.2$/$39.7\pm1.5$/$73.5\pm2.8$ and
$34.9\pm1.3$/-/$11.1\pm0.4$ for $\psi(3770)\to\gamma\chi_{c0/1/2}$ with
$\chi_{c0/1/2}\to 2(\pi^+\pi^-)$, $K^+K^-\pi^+\pi^-$, $3(\pi^+\pi^-)$ and $K^+K^-$ decays, respectively.
The errors arise from uncertainties in the $\psi(3686)$ resonance parameters,
the integrated luminosity of the $\psi(3770)$ data $\mathcal L_{\psi(3770)}$ and the misidentification rates $\eta$.
In Eq. (\ref{eq:bk_psip}), the number of background events depends on the ratio of
the misidentification rate $\eta$ over the efficiency
$\epsilon_{\psi(3686)}$ of reconstructing $\psi(3686)\to \chi_{cJ}$.
Since $\eta$ and $\epsilon_{\psi(3686)}$ all contain the simulation of $\chi_{cJ}\to LH$,
a possible systematic uncertainty from the simulation of $\chi_{cJ}\to LH$ is canceled here.

\section{Results}
The ratio of the branching fraction for $\psi(3770)\to\gamma\chi_{cJ}$ divided by
the branching fraction for $\psi(3686)\to\gamma\chi_{cJ}$ is
determined channel by channel as
\begin{eqnarray}
\label{eq:Rcj}
R_{cJ}
=
\frac
{{\mathcal B}[\psi(3770)\to\gamma\chi_{cJ}]}
{{\mathcal B}[\psi(3686)\to\gamma\chi_{cJ}]}
=
\frac
{N_{\psi(3770)}\cdot
N^{\rm tot}_{\psi(3686)}\cdot
\epsilon_{\psi(3686)}}
{N_{\psi(3686)}\cdot N^{\rm
tot}_{\psi(3770)}\cdot
\epsilon_{\psi(3770)}},
\end{eqnarray}
where $N_{\psi(3686)}$ and $N_{\psi(3770)}$ are the numbers of $\chi_{cJ}$
observed from the $\psi(3686)$ and $\psi(3770)$ data,
$N^{\rm tot}_{\psi(3686)}$ and $N^{\rm tot}_{\psi(3770)}$ are the
total numbers of $\psi(3686)$ and $\psi(3770)$ decays,
$\epsilon_{\psi(3686)}$ and $\epsilon_{\psi(3770)}$ are the efficiencies
of reconstructing $\psi(3686)$ and $\psi(3770)\to \gamma \chi_{cJ}$
with $\chi_{cJ}\to LH$ estimated by MC simulations, respectively.
Here, $N^{\rm tot}_{\psi(3770)}$ is determined by
$\sigma^{\rm obs}_{\psi(3770)}\cdot \mathcal L_{\psi(3770)}$,
where $\sigma^{\rm obs}_{\psi(3770)}=(7.15\pm0.27\pm0.27)$ nb is the cross section for
$\psi(3770)$ production \cite{plb650_111,prl97_121801,plb652_238} and $\mathcal L_{\psi(3770)}$ is the integrated luminosity
of the $\psi(3770)$ data set \cite{jll_bes3_lum}.

Table \ref{tab:num_data} summarizes the ratios $R_{cJ}$ measured via
the different channels. The results are consistent within statistical
uncertainties. From these measurements, we obtain the statistical-weighted
averages $\bar R_{c0}=(6.89\pm0.28\pm0.65)\%$ and $\bar
R_{c1}=(2.03\pm0.44\pm0.66)\%,$ where the first uncertainty is
statistical and the second systematic.

In the measurements of $\bar R_{c0/1}$, the systematic uncertainty arises
from the uncertainties in the total number (0.81\%) of $\psi(3686)$
decays ($N^{\rm tot}_{\psi(3686)}$ \cite{ref:num_psip}); the
integrated luminosity (1.0\%) of the $\psi(3770)$ data
($L_{\psi(3770)}$ \cite{jll_bes3_lum}); the cross section (5.3\%)
for $\psi(3770)$ ($\sigma^{\rm obs}_{\psi(3770)}$
\cite{plb650_111,prl97_121801,plb652_238}); the photon selection
(1.4\%), assigned based on 1.0\% per photon \cite{bes3_sys_gsel};
the MDC tracking (2.6\%/4.0\%); the particle identification
(2.6\%/4.0\%); the statistical uncertainty (1.0\%) of the efficiency
due to the size of the simulated event sample; the $4C$ kinematic fit (1.0\%), estimated by comparing
the measurements with and without the kinematic fit correction; the
fit to mass spectra (6.4\%/31.5\%), estimated by comparing the
measurements with alternative fit ranges ($\pm20$ MeV/$c^2$),
signal shape (simple Breit-Wigner function) and
background shapes ($\pm1$ order of the polynomial functions);
and the subtraction of $\psi(3686)$ peaking
background (0.5\%/2.0\%). The
efficiencies of the MDC tracking and particle identification for
$K^+$ or $\pi^+$ are examined by the doubly tagged hadronic $D\bar
D$ events. The difference between the efficiencies of data and MC is
assigned as an uncertainty. Then, their effects on $\bar R_{c0/1}$
are estimated to be 2.6\%/4.0\%. Table~\ref{tab:sys_tot} summarizes
these uncertainties. Adding them in quadrature, we obtain the total
systematic uncertainty for $\bar R_{c0/1}$ to be 9.4\%/32.6\%.

\begin{table*}[hbtp]
\begin{center}
\caption{\small
Measured $R_{cJ}$ (\%), where
$N_{\psi(3770)}$ and
$N_{\psi(3686)}$ are
the (peaking background corrected) numbers of $\chi_{cJ}$ observed from the $\psi(3770)$ and $\psi(3686)$ data,
$\epsilon_{\psi(3770)}$ and
$\epsilon_{\psi(3686)}$
are the detection efficiencies (\%).
The uncertainties are statistical only.}
\begin{tabular}{cccc} \hline
$\chi_{cJ}\to LH$ &&$J=0$&$J=1$ \\ \hline
                  &$N_{\psi(3770)}$&$756\pm51$&$80\pm26$\\
                  &$\epsilon_{\psi(3770)}$&$24.1\pm0.2$&$25.7\pm0.2$\\
$2(\pi^+\pi^-)$   &$N_{\psi(3686)}$&$59976\pm318$ &$19712\pm175$\\
                  &$\epsilon_{\psi(3686)}$&$24.9\pm0.2$&$26.5\pm0.2$\\ 
                  &$R_{cJ}$&$6.64\pm0.45$&$2.13\pm0.69$\\ \hline
                  &$N_{\psi(3770)}$&$716\pm54$&$46\pm24$\\
                  &$\epsilon_{\psi(3770)}$&$24.0\pm0.2$&$25.4\pm0.2$\\
$K^+K^-\pi^+\pi^-$&$N_{\psi(3686)}$&$46929\pm240$&$11576\pm115$\\
                  &$\epsilon_{\psi(3686)}$&$23.3\pm0.2$&$24.9\pm0.2$\\ 
                  &$R_{cJ}$&$7.56\pm0.57$&$2.00\pm1.04$\\ \hline
                  &$N_{\psi(3770)}$&$502\pm54$&$76\pm27$\\
                  &$\epsilon_{\psi(3770)}$&$18.5\pm0.2$&$20.0\pm0.2$\\
$3(\pi^+\pi^-)$   &$N_{\psi(3686)}$&$36536\pm237$&$19593\pm153$\\
                  &$\epsilon_{\psi(3686)}$&$18.1\pm0.2$&$19.6\pm0.2$\\ 
                  &$R_{cJ}$&$6.86\pm0.74$&$1.94\pm0.69$\\ \hline
                  &$N_{\psi(3770)}$&$283\pm24$&-\\
                  &$\epsilon_{\psi(3770)}$&$32.5\pm0.2$&-\\
$K^+K^-$          &$N_{\psi(3686)}$&$21452\pm154$ &- \\
                  &$\epsilon_{\psi(3686)}$&$32.1\pm0.2$&- \\ 
                  &$R_{cJ}$&$6.65\pm0.57$&-            \\ \hline
\multicolumn{2}{c}{Averaged $R_{cJ}$}&$6.89\pm0.28$&$2.03\pm0.44$            \\ \hline
\end{tabular}
\label{tab:num_data}
\end{center}
\end{table*}

\begin{table*}[hbtp]
\begin{center}
\caption{\small
Systematic uncertainties (\%) in the measurements of $\bar R_{cJ}$.}
\begin{tabular}{lcc} \hline
      &$\bar R_{c0}$     &$\bar R_{c1}$     \\ \hline
$N^{\rm tot}_{\psi(3686)}$ \cite{ref:num_psip}&0.81&0.81 \\
$\sigma^{\rm obs}_{\psi(3770)}$ \cite{plb650_111,prl97_121801,plb652_238}&5.3 &5.3   \\
$L_{\psi(3770)}$ \cite{jll_bes3_lum}     &1.0 &1.0  \\
MC statistics        &1.0 &1.0  \\ 
Photon selection         &1.4 &1.4  \\
MDC tracking             &2.6 &4.0  \\
Particle identification  &2.6 &4.0  \\
$4C$ kinematic fit        &1.0 & 1.0  \\
Fit to mass spectra  &6.4 &31.5  \\ 
Background subtraction&0.5 & 2.0  \\
 \hline
Total                &9.4 &32.6  \\ \hline
\end{tabular}
\label{tab:sys_tot}
\end{center}
\end{table*}

Multiplying $\bar R_{cJ}$ by the branching fraction ${\mathcal B}[\psi(3686)\to\gamma\chi_{cJ}]$
(and the total width $\Gamma^{\rm tot}_{\psi(3770)}$) taken from the PDG \cite{pdg2014},
we obtain the branching fractions (and the partial widths) for $\psi(3770)\to\gamma\chi_{cJ}$,
which are summarized in Table~\ref{tab:measure_pwcj},
where the first uncertainty is statistical and the second systematic.
In the measurement of ${\mathcal B}[\psi(3770)\to\gamma\chi_{cJ}]$
(and $\Gamma[\psi(3770)\to\gamma\chi_{cJ}]$), the systematic uncertainty
arises from the uncertainties of $\bar R_{c0/1}$ and
the uncertainties of ${\mathcal B}[\psi(3686)\to\gamma\chi_{c0/1}]$ of 2.7/3.2\%
(and the uncertainty of $\Gamma^{\rm tot}_{\psi(3770)}$ of 3.7\%).

\newpage
\section{Summary}

In summary, by analyzing 2.92 fb$^{-1}$ of $e^+e^-$ collision data taken at
$\sqrt s=
3.773~\rm GeV$ and $106.41\times 10^{6}$ $\psi(3686)$ decays taken
at $\sqrt s= 3.686~\rm GeV$ with the BESIII detector at the
BEPCII collider, we measure the branching fraction
${\mathcal B}(\psi(3770)\to\gamma\chi_{c0})=(6.88\pm0.28\pm0.67)\times 10^{-3}$
and the partial width $\Gamma[\psi(3770)\to\gamma\chi_{c0}]=(187\pm8\pm19)~\rm keV$.
These are obtained by first measuring the ratio with respect to the
accurately known branching fraction for $\psi(3686)\to\gamma\chi_{cJ}$ decays.
Our results are, thereby, not influenced by the uncertainties in the branching
fractions of $\chi_{cJ}$ decays to light hadrons as done in Ref.~\cite{prd74_031106}.
The branching fraction and partial width for $\psi(3770) \to \gamma \chi_{c1}$
measured in this work are consistent with our previous measurement \cite{bes3_psipp_gchi1}
within errors.
Table~\ref{tab:measure_pwcj} compares the
$\Gamma[\psi(3770)\to\gamma\chi_{c0/1}]$ measured at BESIII with those
measured by CLEO~\cite{prd74_031106,prl96_182002}~\footnote{
The CLEO measurements were based on
the total width $\Gamma^{\rm tot}_{\psi(3770)}=(23.6\pm2.7)$
MeV \cite{pdg2004} and the cross section
$\sigma^{\rm obs}_{\psi(3770)\to D\bar D}=(6.39\pm0.10^{+0.17}_{-0.08})$ nb
for determining the total number of $\psi(3770)$ decays,
where the $\psi(3770)\to$ non-$D\bar D$ decays were neglected.
In addition, CLEO~\cite{prd74_031106} cited
${\mathcal B}(\psi(3686)\to \gamma\chi_{c0})=(9.22\pm0.11\pm0.46)\%$ and
${\mathcal B}(\psi(3686)\to \gamma\chi_{c1})=(9.07\pm0.11\pm0.54)\%$
to determine ${\mathcal B}(\psi(3770)\to\gamma\chi_{cJ})$ from Ref.~\cite{prd70_112002}.
Although CLEO determined
the branching fraction of $\psi(3770)\to $ non-$D\bar D$
decays to be $(-3.3\pm1.4^{+4.8}_{-6.6})\%$ and set an
upper limit of 9\% at 90\% confidence level \cite{prl96_092002},
the PDG value for $\Gamma_{\psi(3770)}^{D\bar D}/\Gamma_{\psi(3770)}^{\rm tot}$~\cite{pdg2008}
from four measurements at BESII~\cite{prl97_121801,plb641_145,prd76_122002,plb659_74}
implied the branching fraction for $\psi(3770)\to $ non-$D\bar D$ decays to be $(14.7\pm3.2)\%$.
At present, the PDG value for $\Gamma_{\psi(3770)}^{D\bar D}/\Gamma_{\psi(3770)}^{\rm tot}$
gives the branching fraction of $\psi(3770)\to $ non-$D\bar D$ decays to be
$(7^{+8}_{-9})\%$~\cite{pdg2014}.
Therefore, for better comparisons, we also list the corrected CLEO partial widths
with the same input values as those used in our measurements,
which are
$\sigma^{\rm obs}_{\psi(3770)}=(7.15\pm0.27\pm0.27)$ nb~\cite{plb650_111,prl97_121801,plb652_238},
$\Gamma^{\rm tot}_{\psi(3770)}=(27.2\pm1.0)$ MeV,
${\mathcal B}(\psi(3686)\to \gamma\chi_{c0})=(9.99\pm0.27)\%$ and
${\mathcal B}(\psi(3686)\to \gamma\chi_{c1})=(9.55\pm0.31)\%$~\cite{pdg2014}.
}
and the
theoretical calculations from
Refs.~\cite{prd64_094002,prd44_3562,prd69_094019,prd72_054026,new_cite}. The
partial width $\Gamma[\psi(3770)\to\gamma\chi_{c0}]$ measured at
BESIII is consistent within errors with the one measured by CLEO
with an improved precision. These results underline the fact that
the traditional models \cite{prd44_3562,prd69_094019,prd72_054026}
with a relativistic assumption or a coupled-channel correction
agree quantitatively better with the experimental data
than those \cite{prd64_094002,prd44_3562,prd69_094019,prd72_054026} based upon non-relativistic calculations.
For these traditional models, the non-relativistic calculations clearly
overestimate the partial width $\Gamma[\psi(3770)\to\gamma\chi_{cJ}]$.
The experimental data also support the recent calculation
based on the non-relativistic constituent quark model (NRCQM)~\cite{new_cite}.
Together with further
theoretical developments, our results aim to contribute to a deeper
understanding of the dynamics of charmonium decays above the
open-charm threshold.

\begin{table*}[hbtp]
\begin{center}
\caption{\small
Comparisons of the partial widths for $\psi(3770)\to\gamma\chi_{cJ}$ (in keV),
where ${\mathcal B}$ and $\Gamma$ denote the branching fraction and the
partial width for $\psi(3770)\to\gamma\chi_{cJ}$, respectively.
For the BESIII results, the first uncertainty is statistical and the second systematic.
Detailed explanations about the CLEO results can be found in footnote 1.}
\begin{tabular}{lcc} \hline
Experiments
 &$J=0$&$J=1$\\ \hline
${\mathcal B}^{\rm BESIII}$($\times 10^{-3}$)&$6.88\pm0.28\pm0.67$&$1.94\pm0.42\pm0.64$\\
${\mathcal B}^{\rm BESIII}$($\times 10^{-3}$) \cite{bes3_psipp_gchi1}&--&$2.48\pm0.15\pm0.23$\\ \hline
$\Gamma^{\rm BESIII}$&$187\pm8\pm19$&$53\pm12\pm18$ \\
$\Gamma^{\rm BESIII}$~\cite{bes3_psipp_gchi1}&--&$67.5\pm4.1\pm6.7$ \\
$\Gamma^{\rm CLEO}$~\cite{prd74_031106,prl96_182002} &$172\pm30$&$70\pm17$ \\
$\Gamma_{\rm corrected}^{\rm CLEO}$ &$192\pm24$&$72\pm16$ \\ \hline \hline
Theories & & \\ \hline
Rosner \cite{prd64_094002} (non-relativistic)
&$523\pm12$&$73\pm9$\\ \hline
Ding-Qing-Chao \cite{prd44_3562} & & \\
non-relativistic&312&95\\
relativistic   &199& 72\\ \hline
Eichten-Lane-Quigg \cite{prd69_094019} & & \\
non-relativistic&254&183\\
with coupled channels corrections&225&59\\ \hline
Barnes-Godfrey-Swanson \cite{prd72_054026}& &\\
non-relativistic&403&125\\
   relativistic&213&77\\ \hline
NRCQM \cite{new_cite} & 218 & 70 \\ \hline
\end{tabular}
\label{tab:measure_pwcj}
\end{center}
\end{table*}

\section{Acknowledgements}
The BESIII collaboration thanks the staff of BEPCII and the IHEP computing
center for their strong support. This work is supported in part by National
Key Basic Research Program of China under Contract Nos. 2009CB825204 and 2015CB856700;
National Natural Science Foundation of China (NSFC) under Contracts Nos. 10935007,
11125525, 11235011, 11305180, 11322544, 11335008, 11425524; the Chinese Academy of
Sciences (CAS) Large-Scale Scientific Facility Program; Joint Large-Scale
Scientific Facility Funds of the NSFC and CAS under Contracts Nos. 11179007,
U1232201, U1332201; CAS under Contracts Nos. KJCX2-YW-N29, KJCX2-YW-N45;
100 Talents Program of CAS; the CAS Center for Excellence in Particle Physics
(CCEPP); INPAC and Shanghai Key Laboratory for Particle
Physics and Cosmology; German Research Foundation DFG under Contract No.
Collaborative Research Center CRC-1044; Istituto Nazionale di Fisica Nucleare,
Italy; Ministry of Development of Turkey under Contract No. DPT2006K-120470;
Russian Foundation for Basic Research under Contract No. 14-07-91152; U. S.
Department of Energy under Contracts Nos. DE-FG02-04ER41291, DE-FG02-05ER41374,
DE-FG02-94ER40823, DESC0010118; U.S. National Science Foundation; University of
Groningen (RuG) and the Helmholtzzentrum fuer Schwerionenforschung GmbH (GSI),
Darmstadt; WCU Program of National Research Foundation of Korea under Contract
No. R32-2008-000-10155-0.

\end{document}

%% file: BESIII_authors.tex
\author{
  \begin{small}
    \begin{center}
      M.~Ablikim$^{1}$, M.~N.~Achasov$^{9,f}$, X.~C.~Ai$^{1}$,
      O.~Albayrak$^{5}$, M.~Albrecht$^{4}$, D.~J.~Ambrose$^{44}$,
      A.~Amoroso$^{49A,49C}$, F.~F.~An$^{1}$, Q.~An$^{46,a}$,
      J.~Z.~Bai$^{1}$
% ...
      \\
      \vspace{0.2cm}
      (BESIII Collaboration)\\
      \vspace{0.2cm} {\it
        $^{1}$ Institute of High Energy Physics, Beijing 100049, People's Republic of China\\
% ...
$^{4}$ Bochum Ruhr-University, D-44780 Bochum, Germany\\
        $^{5}$ Carnegie Mellon University, Pittsburgh, Pennsylvania 15213, USA\\
% ...
$^{9}$ G.I. Budker Institute of Nuclear Physics SB RAS (BINP), Novosibirsk 630090, Russia\\
% ...
$^{44}$ University of Rochester, Rochester, New York 14627, USA\\
% ...
$^{46}$ University of Science and Technology of China, Hefei 230026, People's Republic of China\\
% ...
$^{49}$ (A)University of Turin, I-10125, Turin, Italy; (B)University of Eastern Piedmont, I-15121, Alessandria, Italy; (C)INFN, I-10125, Turin, Italy\\
% ...
$^{a}$ Also at State Key Laboratory of Particle Detection and Electronics, Beijing 100049, Hefei 230026, People's Republic of China\\
% ...
$^{f}$ Also at the Novosibirsk State University, Novosibirsk, 630090, Russia\\
% ...
      }\end{center}
    \vspace{0.4cm}
  \end{small}
}